\documentclass[twocolumn,prl,a4paper,superscriptaddress,showpacs,floatfix]{revtex4}
\usepackage{amsmath}
\usepackage{graphicx}

\newcommand{\Uvec}{\mathbf{U}}

\newcommand{\wsr}{\gamma}
\newcommand{\wsrm}{\wsr_m}
\newcommand{\Um}{U_m}

\newcommand{\gradpm}{(\nabla p){}_m}
\newcommand{\paper}{Letter}
\newcommand{\KC}{Kozeny-Carman}
\newcommand{\latin}[1]{\emph{#1}}
\newcommand{\etal}{\latin{et al.}}
\newcommand{\ie}{\latin{i.$\,$e.}}

\newcommand{\KCfac}{c_0}
\newcommand{\wsrprefac}{\alpha}

\begin{document}

\title{The wall shear rate distribution for flow in random sphere packings}

\author{Patrick B. Warren}
\affiliation{Unilever R\&D Port Sunlight, Bebington, Wirral, CH63 3JW, UK.}

\author{Frantisek Stepanek}
\affiliation{Chemical Engineering and Chemical Technology, Imperial
College, London, SW7 2AZ, UK.}

\date{October 22nd, 2007 --- revised submission of LF11477 Warren}

\begin{abstract}
The wall shear rate distribution $P(\wsr)$ is investigated for
pressure-driven Stokes flow through random arrangements of spheres at
packing fractions $0.1\le\phi\le0.64$.  For dense packings, $P(\wsr)$
is monotonic and approximately exponential.  As $\phi\to0.1$,
$P(\wsr)$ picks up additional structure which corresponds to the flow
around isolated spheres, for which an exact result can be obtained.  A
simple expression for the mean wall shear rate is presented, based on a
force-balance argument.
\end{abstract}

% 47.56.+r Flows through porous media 
% 47.15.G- Low-Reynolds-number (creeping) flows


\maketitle

\begin{figure}
\includegraphics{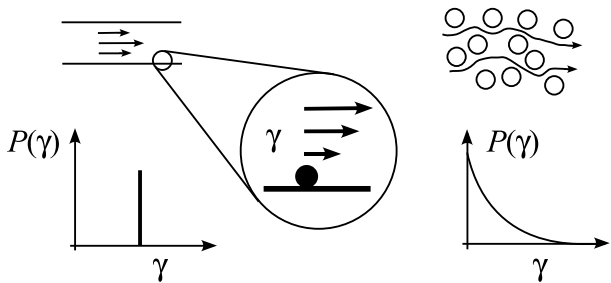}
\caption[]{The flow in the controlled geometry of a flow cell gives
rise to a uniform wall shear rate (left), whereas the flow in a porous
material gives rise to a \emph{distribution} of wall shear rates
(right).  It is the wall shear rate $\wsr$ that governs the
deposition and detachment of particles (inset).\label{fig:flow2d}}
\end{figure}

The wall shear rate $\wsr$ is the rate at which the tangential
velocity of a fluid vanishes on approaching a wall.  It determines the
hydrodynamic forces acting on a particle adjacent to the wall, and is
therefore a key quantity governing deposition, retention, and
detachment \cite{vandeven}.  For example, $\wsr$ is an important
factor which determines whether colloidal particles can become
attached to a surface by specific ligand binding \cite{CTH-Pangu}.
Experimentally, such processes are often examined using flow cells
with well-controlled hydrodynamics for which the wall shear rate is
known.  For flow through a porous material though, one can expect a
\emph{distribution} of wall shear rates $P(\wsr)$.  This is
illustrated in Fig.~\ref{fig:flow2d}.  A crucial issue for realistic
situations, such as deep bed filtration \cite{TGMvdV-RydeM} or
particulate soil detergency in fabric cleaning \cite{Kissa-Carroll},
is therefore to characterise the wall shear rate distribution
$P(\wsr)$ for flow in more complex pore spaces.  This is also a
problem of generic interest in the growing field of statistical
microhydrodynamics.  Previously, only $P(\wsr)$ for flow in
two-dimensional channels with (fractally) rough walls has been
investigated \cite{AndradeX}.

In this \paper, $P(\wsr)$ and its relation to the mean fluid
velocity $\Um$ is investigated for pressure-driven Stokes flow
in random sphere packings at packing fractions in
the range $0.1\le\phi\le0.64$.  The relationship between $P(\wsr)$
and $\Um$ is of paramount importance for applications since it is
very difficult to access $P(\wsr)$ experimentally (it either has to
be done by detailed resolution of the flow field, or indirectly by
looking at the behaviour of particulate tracers), but determination of
$\Um$ is much easier.

We generated sphere packings in periodic simulation boxes for
packing fractions in the range $0.1\le\phi\le0.64$ by a
Monte-Carlo algorithm \cite{SLL}.  The highest
packing fraction corresponds to the usual random close-packed limit.
Whilst the lower packing fractions are mechanically unstable, they
provide a useful interpolation between isolated spheres and packed
beds.  We also generated a slightly looser packing of touching spheres
at $\phi\approx0.622$ by a sequential deposition algorithm \cite{CTA}.
This latter geometry is not periodic in the $z$-direction (the
deposition direction), but we have found that the bulk properties can
be determined by extrapolation.

For the flow-field calculations we use a standard lattice Boltzmann
(LB) methodology which is now well-developed for this class of
problems \cite{Ladd12, HL, KoponenX, MAKHTX, MKDBX, vdHBK, PLM,
note-code}.  As already mentioned, we solve the Stokes equations and
thus operate at strictly zero Reynolds number.  The spheres are held
stationary and flow is generated in the pore space by applying a
uniform body force corresponding to a mean pressure gradient $\gradpm$
in the $x$-, $y$- or $z$-directions.  The hydrodynamic forces exerted
on wall lattice nodes are easily found in LB.  For each wall node one
can trivially determine the tangential force component since the
corresponding sphere centre is known.  The local wall shear rate is
then given by the magnitude of the tangential component divided by the
viscosity.  In this way we obtain a large set of wall shear rates from
which we reconstruct $P(\wsr)$ \cite{note-weight}.  We also measure
the mean volumetric (superficial) fluid velocity $\Um$.

\begin{figure}
\includegraphics{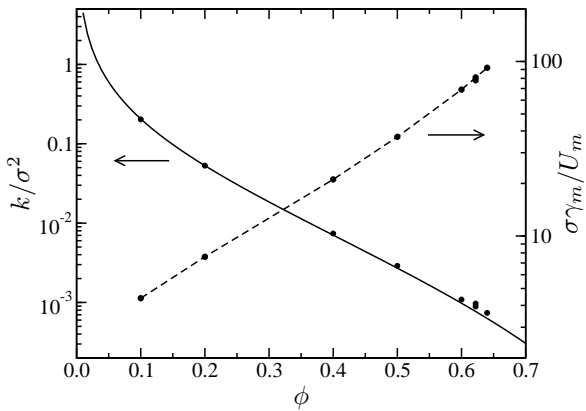}
\caption[]{Dimensionless permeability and mean wall shear rate as a
function of packing fraction, from Table \ref{tab:res}.  The solid
line is Eq.~(31) from van der Hoef \etal\ \cite{vdHBK} which is
claimed to be accurate to within 3\%.  The dashed line for the mean wall
shear rate data is a guide to the eye. Error bars are smaller than the
symbols.\label{fig:res}}
\end{figure}

We first discuss our results for the permeability, $k$, since this
underpins our analysis of $P(\wsr)$.  It is defined via Darcy's law,
$\Um=(k/\eta)\gradpm$, where $\eta$ is the viscosity.  Our results,
expressed in dimensionless terms using $k/\sigma^2$, are shown as a
function of packing fraction in Table~\ref{tab:res} and
Fig.~\ref{fig:res}.  Generally speaking, the permeability falls
dramatically with increasing packing fraction.  For $\phi\le0.5$ our
results are in excellent agreement with previous work by Ladd
\cite{JCP23} and van der Hoef \etal\ \cite{vdHBK}.  For $\phi\ge0.6$
our results are $\approx10\%$ higher than the accurate results
obtained recently by van der Hoef \etal\ \cite{vdHBK}, although we are
in agreement with some previous studies \cite{HL, TWP}.  This may
reflect subtle differences in the way the sphere packings are
constructed.  The sequential deposition packing at $\phi\approx0.622$
fits nicely into the series.  In this case the permeability is
in principle different parallel to and perpendicular to the deposition
direction.  We find though that the difference is certainly less than
10\%, in agreement with Coelho \etal\ \cite{CTA}.

An oft-used correlation is the \KC\ relation,
\begin{equation}
k = {(1-\phi)^3}\!/{\KCfac s^2},\label{eq:kc}
\end{equation}
where $s=6\phi/\sigma$ is the specific surface area of spheres (in any
arrangement) and the numerical factor $\KCfac\approx4$--$5$
\cite{Carman, HL, MKDBX, TWP}.  We find this does indeed capture the
behaviour of the permeability quite well for intermediate to high
packing fractions (Table~\ref{tab:res}).  Interestingly, for
$\phi\ge0.2$ we noticed our data can be accurately fit by
$\log(k/\sigma^2) = A + B \phi$ with $A = -1.04(6)$ and $B = -9.6(1)$,
reminiscent of what has been found for fibrous beds \cite{KoponenX}.

Now we turn to the mean wall shear rate, defined via $\wsrm =
\int_0^\infty\! d\wsr\, \wsr\, P(\wsr)$.  For Stokes flow,
$\wsrm$ is strictly proportional to $\Um$, so that
$\sigma\wsrm/\Um$ is a convenient way to express the mean wall
shear rate in dimensionless terms, shown in
Table~\ref{tab:res} and Fig.~\ref{fig:res}.  We see
that $\sigma\wsrm/\Um$ grows dramatically with packing fraction,
similar to the inverse of $k/\sigma^2$.  

This behaviour can be understood by the following force-balance
argument.  The force per unit volume acting on the fluid due to the
mean pressure gradient is $(1-\phi)\gradpm$.  In steady state this
must balance the integrated wall stress, thus the mean wall stress is
\emph{exactly} $(1-\phi)\gradpm/s$ where $s$ is the specific surface
area.  If we now \emph{approximate} the mean wall stress by $\eta\wsrm$,
use Darcy's law to replace $\gradpm$ by $\Um$, and substitute
$s=6\phi/\sigma$, we get
\begin{equation}
\wsrm
= \wsrprefac{(1-\phi)\sigma\Um}/{(6\phi k)}.\label{eq:wsr}
\end{equation}
We have introduced a prefactor $\wsrprefac$ to capture the approximate
nature of this expression.  From our data we find that
$\wsrprefac\approx0.6$--$0.8$ is very insensitive to packing fraction,
as shown in Table~\ref{tab:res} (we can rationalise this value of
$\alpha$ by arguing that, on average, $2/3$ of the wall stress lies in
the wall tangent plane).  Eq.~\eqref{eq:wsr} explains the approximate
inverse relationship between $\sigma\wsrm/\Um$ and $k/\sigma^2$.
Incidentally, in a parallel-sided capillary of arbitrary cross section,
the flow is laminar and parallel to the walls.  In this case the mean wall
stress is exactly $\eta\wsrm$ and Eq.~\eqref{eq:wsr} is exact with
$\wsrprefac\equiv1$.  Our LB methodology is constructed to retain this
exact result, provided the capillary axis is aligned with a grid
axis.

\begin{table}
\begin{ruledtabular}
\begin{tabular}{lllllll}
$\phi$
 && \multicolumn{1}{c}{$k/\sigma^2\times 10^3$}  & \multicolumn{1}{c}{$\KCfac$}
 && \multicolumn{1}{c}{$\sigma\wsrm/\Um$} & \multicolumn{1}{c}{$\wsrprefac$} 
\\[2pt]
\hline
0.1          && 203(8)              
 &           10.0(4) && \phantom{0}4.4(2) & 0.60(2) \\
0.2          && \phantom{0}53(2)    
 & \phantom{0}6.7(3) && \phantom{0}7.6(3) & 0.60(2) \\
0.4          && \phantom{00}7.4(3)  
 & \phantom{0}5.1(2) && 21.1(5)           & 0.62(3) \\
0.5          && \phantom{00}2.9(1)  
 & \phantom{0}4.8(2) && 37(1)\phantom{.0} & 0.64(2) \\
0.6          && \phantom{00}1.09(6) 
 & \phantom{0}4.5(2) && 69(2)\phantom{.0} & 0.68(4) \\
0.622 ($z$)  && \phantom{00}0.97(7) 
 & \phantom{0}4.0(3) && 81(7)\phantom{.0} & 0.78(6) \\
0.622 ($xy$) && \phantom{00}0.89(6) 
 & \phantom{0}4.4(3) && 78(6)\phantom{.0} & 0.69(5) \\
0.64         && \phantom{00}0.74(4) 
 & \phantom{0}4.3(2) && 92(4)\phantom{.0} & 0.73(4) \\
\end{tabular}
\end{ruledtabular}
\caption[?]{Dimensionless permeability $k/\sigma^2$ and mean wall
shear rate $\sigma\wsrm/\Um$ as a function of packing fraction $\phi$:
$\KCfac$ is the \KC\ factor in Eq.~\eqref{eq:kc} and $\wsrprefac$ is
the prefactor in the force-balance expression in Eq.~\eqref{eq:wsr}.
For the sequential deposition sample ($\phi\approx0.622$), results are
given parallel and perpendicular to the deposition direction ($z$).  A
figure in brackets is an estimate of the error in the final
digit \cite{note-code}.\label{tab:res}}
\end{table}

\begin{figure}
\includegraphics{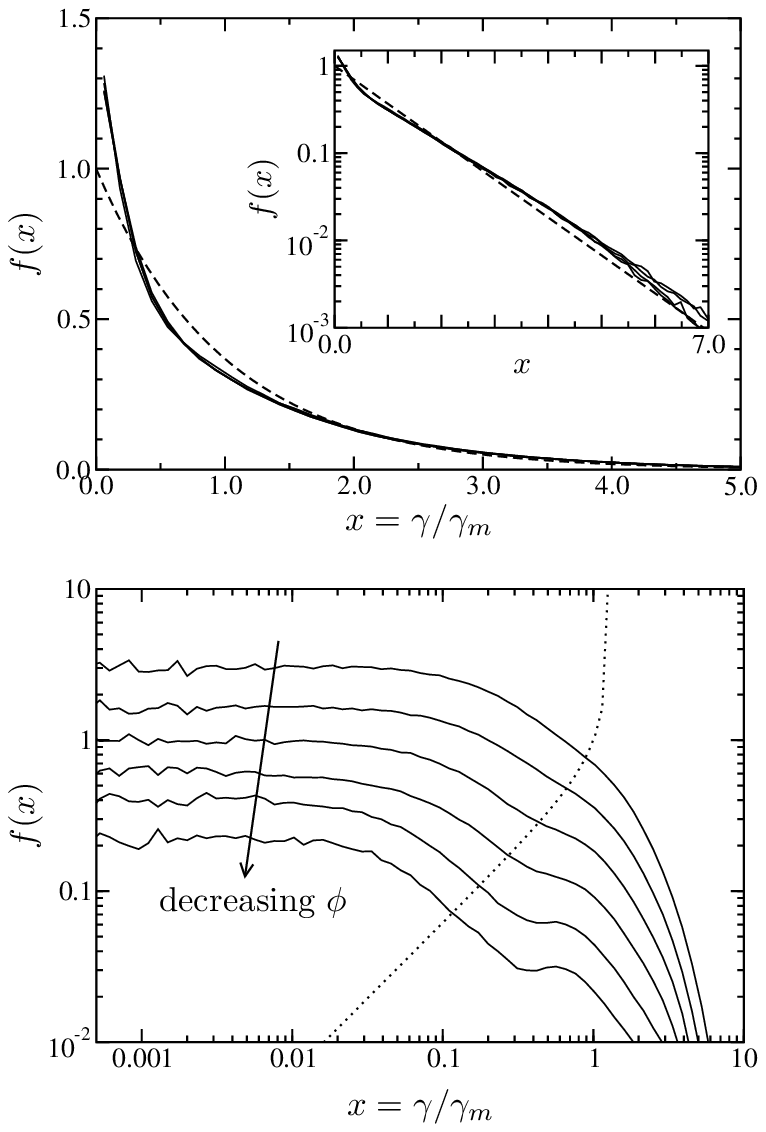}
\caption[]{The upper plot shows the wall shear rate distributions for
all data sets with $\phi\ge0.6$.  The dashed line is $f(x)=e^{-x}$.
The lower plot shows the same for the six periodic packings with
$0.1\le\phi\le0.64$; the curves are displaced for clarity. The dotted
line is the exact result in Eq.~\eqref{eq:stokes} for Stokes flow
around an isolated sphere.\label{fig:distrib}}
\end{figure}

Finally we turn to the wall shear rate distribution, which we report
in terms of $x=\wsr/\wsrm$ and $f(x)$ defined such that
$P(\wsr)=(1/\wsrm)\,f(\wsr/\wsrm)$.  At packing fractions
$\phi\ge0.6$, $f(x)$ is monotonic and quite well approximated by an
exponential (Fig.~\ref{fig:distrib}, upper plot).  It is interesting
to note that a similar exponential distribution is found for the local
flow speeds although in this case a peak at zero is to be expected
given the large volume of pore space immediately adjacent to the
sphere surfaces \cite{MKDBX, MAKHTX}.  We will return to the small $x$
behaviour of $f(x)$ in a moment.

As the packing fraction is reduced, a hump appears in $f(x)$ at around
$x=0.5$--$0.6$ (Fig.~\ref{fig:distrib}, lower plot).  This feature
seems to be associated with the transition from channel flow at high
packing fractions towards flow around individual spheres at lower
packing fractions.  This interpretation is supported by the exact
result which can be obtained for $P(\wsr)$ from the Stokes solution
for flow around a sphere, as we now discuss.

A remarkable feature of Stokes flow around a sphere is that the wall
stress has the same vectorial value $3\eta\Uvec/\sigma$ at all points
on the sphere surface, where $\Uvec$ is the flow velocity at infinity
\cite{Batchelor}.  If we project this into the wall tangent plane, we
obtain the local wall shear rate $\wsr={(3\Um\sin\theta)}/{\sigma}$,
where $\theta$ is the angle between the wall normal and the direction
of the flow field at infinity, and $\Um\equiv|\Uvec|$.  The mean wall
shear rate is then given by $\sigma\wsrm/\Um = \int_0^\pi
{(3/2)\sin^2\!\theta\,d\theta} = {3\pi}/{4}\approx2.356$.  It follows
that $x = {\wsr}/{\wsrm}=({4}/{\pi})\sin\theta$, and from $f(x)\,dx =
(1/2)\sin\theta\,d\theta$ (\ie\ the area measure \cite{note-weight}),
\begin{equation}
f(x)=\frac{\pi x/4}{\sqrt{(4/\pi)^2-x^2}},\quad 0\le x\le{4}/{\pi}.
\label{eq:stokes}
\end{equation}
This is the desired exact result for the wall shear rate distribution
for Stokes flow around an isolated sphere, shown as the dotted line in
the lower plot of Fig.~\ref{fig:distrib}.  It diverges as $x \to
4/\pi\approx1.273$, corresponding to $\theta \to \pi/2$ where the wall
shear rate is maximal.  This behaviour is, we believe, responsible for
the hump that appears in $f(x)$ at low packing fractions.  The fact
that there is still a significant difference between
Eq.~\eqref{eq:stokes} and $f(x)$ for
$\phi=0.1$ should not be too surprising given the long range nature
of hydrodynamic interactions.  We see this also in $k$ and $\wsrm$
which are, respectively, a factor $\approx2.76$ smaller and a factor
$\approx1.9$ higher, than the corresponding isolated sphere limits
(\ie\ $k/\sigma^2 = 1/(18\phi)$ \cite{MTB, JCP23} and
$\sigma\wsrm/\Um=3\pi/4$ derived above).  In fact the permeability
data from Ladd suggests that the isolated sphere result is approached
only very slowly as $\phi\to0$ \cite{JCP23}.

Now we return to the small $x$ behaviour of $f(x)$.  Clearly, for any
sphere, the local wall shear rate has to vanish at least at one point
on the sphere surface---this is a consequence of the so-called `hairy
ball theorem' \cite{note-hairyball}.  Thus it is not at first sight
surprising that $f(x)$ goes to a plateau as $x\to0$
(Fig.~\ref{fig:distrib}, lower plot).  However, Eq.~\eqref{eq:stokes}
has the property that $f(x)\sim x$ as $x\to0$ arising from the
stagnation points at $\theta=(0,\pi)$.  This behaviour might be
expected to be generic for low packing fractions where stagnation
points are common.  In contrast, for dense sphere packings the flow is
more channel-like and stagnation points are rare.  In this case the
wall shear rate vanishes, \latin{inter alia}, at all the contact
points between spheres.  Analysis of pressure-driven flow in the
vicinity of a contact point using the Reynolds lubrication equation
\cite{Oron} suggests $f(x)\sim x^{\delta}$ for $x\to0$ where $\delta=
{(4-\surd10)} / {(\surd10-2)} \approx 0.72$.  It is therefore rather
surprising that, independent of packing fraction, a \emph{plateau}
rather than a power law is observed for $f(x)$ as $x\to0$.

One possible reason for this is that long-range flow field
inhomogeneities (on length scales $\agt\sigma$) wash out the expected
behaviour and replace the power law by a plateau.  We investigated
this possibility by constructing an individual $f(x)$ for each sphere,
then averaging over all the spheres in a sample.  This should remove
the effects of long-range flow field inhomogeneities.  We find though
there is little change in $f(x)$; the hump at low $\phi$ becomes
somewhat more pronounced but the plateau remains in all cases.  At the
same time we also examined the hydrodynamic forces acting on
individual spheres.  We found that these have a relatively narrow
distribution (approximately Gaussian, with a standard deviation
20--30\% of the mean) indicating that the flow field on length scales
$\agt\sigma$ is rather homogeneous.  We conclude that long-range flow
field inhomogeneities are unlikely to be important.  Instead, the
implication is that the shape of $f(x)$, and in particular the plateau
at $x\to0$, is mostly controlled by the local pore geometry.  The
important message seems to be that using highly idealised situations,
such as the Stokes solution for flow around an isolated sphere or
lubrication theory in the vicinity of a contact point, may give
qualitatively misleading results when it comes to inferring the
overall statistical properties.

To summarise, for applications Eq.~\eqref{eq:wsr} provides the key
link between the mean wall shear rate $\wsrm$ and the mean fluid
velocity $\Um$.  If necessary the Darcy permeability can be estimated
from the \KC\ relation in Eq.~\eqref{eq:kc}.  Knowledge of $\wsrm$ is
then sufficient to determine the whole wall shear rate distribution,
if the latter is assumed to be exponential, \ie\ $P(\wsr) \approx
(1/\wsrm) \exp(-\wsr/\wsrm)$.  More generally, our investigation
demonstrates how direct numerical calculation of the statistical
properties of microhydrodynamic flows can complement exact solutions
for simplified geometries, to gain new insights.

We thank Theo van de Ven for helpful discussions, and the Unilever
Centre for Molecular Science Informatics in the University of
Cambridge for access to the computational resources of the `CamGrid'
network.

%\bibliography{wsnotes,wallshear}

\begin{thebibliography}{23}
\bibitem[{not({\natexlab{a}})}]{note-code}
\bibinfo{note}{We use a 15-velocity LB scheme with link-bounce-back, similar to
  that developed by Ladd \cite{Ladd12}. To reduce discretisation errors we use
  a grid refinement procedure similar to van der Hoef \etal\ \cite{vdHBK}.
  Typically 5 sphere packings of size $(5\sigma)^3$ times 3 flow directions
  were used for each $\phi$. The errors arising from sample averaging and grid
  refinement are estimated to be both of the order 2--3\%.}
\bibitem[{not({\natexlab{b}})}]{note-weight}
\bibinfo{note}{An area measure is used in computing the statistical properties,
  in other words, equal weights are assigned to equal wall areas. This is by no
  means the only measure that could be contemplated, although we leave
  exploration of this aspect to future work.}
\end{thebibliography}

\end{document}